\documentstyle[preprint,aps]{revtex}

\newcommand{\mrm}[1]{\mbox{\rm #1}}
\newcommand{\beq}{\begin{equation}}
\newcommand{\eeq}{\end{equation}}
\newcommand{\nn}{\nonumber}
\newcommand{\bea}{\begin{eqnarray}}
\newcommand{\eea}{\end{eqnarray}}

\newcommand{\rfn}[1]{(\ref{#1})}
\newcommand{\Eq}[1]{Eq.~(\ref{#1})}

\newcommand{\ea}{{ et al.}}

\newcommand{\np}[1]{{ Nucl. Phys. }{\bf #1}}
\newcommand{\plet}[1]{{ Phys. Lett. }{\bf #1}}
\newcommand{\pr}[1]{{ Phys. Rev. }{\bf #1}}
\newcommand{\prlet}[1]{{ Phys. Rev. Lett. }{\bf #1}}
\newcommand{\zp}[1]{{ Z. Phys. }{\bf #1}}

\newcommand{\ptp}[1]{{ Prog. Theor. Phys. }{\bf #1}} 
\newcommand{\arnps}[1]{{ Ann. Rev. Nucl. Part. Sci. }{\bf #1}}

\def\lsim{\mathrel{\vcenter{\hbox{$<$}\nointerlineskip\hbox{$\sim$}}}}
\def\gsim{\mathrel{\vcenter{\hbox{$>$}\nointerlineskip\hbox{$\sim$}}}}

\begin{document}

\preprint{FTUV/97-61}

\draft

\title{CP Asymmetries in $B^0$ Decays in the Left-Right Model}

\author{G. Barenboim$^1$, J. Bernab\'eu$^1$  and M. Raidal$^{1,2}$}

\address{
        $^1$Department of Theoretical Physics, University of Valencia,
        E-46100 Burjassot, Valencia, Spain \\
        $^2$KBFI, Academy of Estonia, R\"avala 10, EE-0001 Tallinn, Estonia}
\date{\today}
\maketitle
\begin{abstract}
We study the time dependent CP asymmetries in $B^0_{d,s}$ decays in 
the left-right model  with spontaneous breakdown of CP. 
Due to the new contributions to $B^0$-$\bar B^0$ mixing, 
the CP asymmetries can be substantially modified. 
Moreover, there can be significant new contributions to the $B$-meson decay 
amplitudes from the magnetic penguins.
Most promising for detection of the 
new physics in the planned $B$ factories is that the CP asymmetries 
in the decays $B\to J/\psi K_S$ and $B\to \phi K_S$ which are supposed to be 
equal in the standard model can differ 
significantly in this class of models independently of the results in 
the measurements of $B\to X_s \gamma.$
\end{abstract}


CP violation, currently observed only in the neutral kaon system, 
 is one of the least tested aspects of Nature.
The standard model (SM) has specific predictions on the size as well as on
the patterns of CP violation in $B$ meson decays \cite{buras}
which, if disproved in the future $B$ factories, 
would signal the existence of new physics \cite{quinn}.   
In $B^0$  decays  new physics can possibly contribute 
 to the $B^0_q$-$\bar B^0_q$ ($q=d,s$) mixing  
as well as  to the decay amplitudes.  The effect of the 
new physics in the mixing  is universal, i.e., the time dependent 
rate asymmetries between $B^0_q$ and $\bar B^0_q$ in all their decays to 
the common CP eigenstates receive the same contribution. 
On the other hand, the effects of new physics in the
decay amplitudes are non-universal and can show up in the comparison of
the CP asymmetries in different decay modes \cite{gross,susy}.

 In this Letter we analyze the CP asymmetries in $B^0$ decays
in the $\mrm{SU}(2)_R\times \mrm{SU}(2)_L\times \mrm{U}(1)_{B-L}$ 
left-right symmetric model (LRSM) \cite{lr} with spontaneous
breakdown of CP \cite{lee,cpk,cpb}. 
Indeed, in such a model with spontaneous parity violation 
it is natural to consider also CP as a spontaneously broken symmetry.
We  show that with the present constraints on the parameters of
the right-handed sector the new contribution to $B$ meson mixing 
can be large and time dependent CP asymmetries can vary from -1 to 1 
in both $B_d^0$ and $B_s^0$ systems. In addition, due to the 
new penguins contributing to the flavor changing decay $b\to s\bar s s$
the CP asymmetries in $B\to J/\psi K_S$ and $B\to K_S\phi$ which 
with high accuracy  measure the same unitary triangle angle, $\beta$,
 in the SM
may  differ from each other almost by unity in the LRSM even in the case
in which the measurements in  $B\to X_s \gamma$ correspond exactly to the 
SM predictions. These two effects are complementary, while the former one
is dominated by the new heavy particle exchange,  the latter one  
is due to the left-right mixing.

The Higgs sector of the LRSM contains a bidoublet 
$\Phi(\frac{1}{2}$, $\frac{1}{2}$, 0) and two triplets, 
 $\Delta_L$(1, 0, 2) and $\Delta_R$(0, 1, 2).
In order to have parity as a spontaneously broken symmetry,
a discrete left-right symmetry, 
$\Psi_{iL} \leftrightarrow \Psi_{iR},$
$\Delta_L  \leftrightarrow \Delta_R, $
$ \Phi  \leftrightarrow \Phi^\dagger,$
should be  imposed.  
After spontaneous  symmetry breaking,  the vacuum expectation values (vev) of
the neutral components of $\Phi,$ $k_1$ and $k_2  e^{i \omega}$
give masses to  the quarks and  left-handed gauge bosons.
The phase $\omega$ which is the relative phase between the 
vev's is the only source of  CP-violation in our model. 
The left- and right-handed Cabbibo-Kobayashi-Maskawa 
(CKM) matrices $V_L$ and $V_R,$ respectively, 
are related as
$|V_L|=|V_R|$,
due to the discrete left-right symmetry. 
They contain all together six CP phases
which are related to $\omega.$
In the following it would be convenient to think about $V_L$ as the 
SM CKM matrix and to shift all the phases but one to $V_R.$  
The charged current Lagrangian in the LRSM  is given by 
${\cal L}_{cc} = g/\sqrt{2} \overline{u} ( \cos\xi V_L 
\gamma^\mu P_L - e^{i \omega} \sin\xi V_R \gamma^\mu P_R ) d \;
W_{1\mu} +g/\sqrt{2} \overline{u} ( 
e^{-i\omega} \sin\xi V_L 
\gamma^\mu P_L + \cos\xi V_R \gamma^\mu P_R ) d \;
W_{2\mu}  +  \mbox{H.c.},$
where $P_{L,R} \equiv (1 \mp \gamma_5)/2,$ 
$W_1,$ $W_2$ are the charged vector boson fields with the masses
$M_1,$ $M_2,$ respectively, and $\xi$ denotes their mixing.
The appearance of $\omega$ in the charged current Lagrangian
 is pure convention
since it can be removed to $V_R.$ The most stringent lower bound on
$W_2$ mass, $M_2\gsim 1.6$ TeV, is derived from the $K_L$-$K_S$
mass difference \cite{beall}. The experimental upper bound on the mixing 
angle $\xi$ depends on the phase  $\omega.$ For small
phases it is 
$\xi\lsim 0.0025$ while for large phases $\xi\lsim 0.013$ \cite{lang}.
All these results are subject of large hadronic uncertainties.
The best limit on $\xi$,  
free of these uncertainties, arises from the muon decay data
and is $\xi\lsim 0.033$ \cite{meie}. However, for our numerical evaluations 
we use the appropriate stringent bounds from Ref. \cite{lang}.
There are two neutral flavor changing Higgs bosons in the model whose
masses are constrained as $M_H\gsim 12$ TeV \cite{cpb,comment}.

CP violation in $B^0$ decays  
takes place 
 due to the interference between mixing and decay. The corresponding
CP asymmetry depends on the parameter $\lambda$
defined as \cite{quinn}
\bea
\lambda=\left( \sqrt{\frac{M_{12}^*-\frac{i}{2}\Gamma_{12}^*}
{M_{12}-\frac{i}{2}\Gamma_{12}}}\right)\frac{\bar A}{A}=e^{-2i\phi_M}
\frac{\bar A}{A}\,,
\label{lambda}
\eea
where $A$ and $\bar A$ are the amplitudes of $B^0$ and $\bar B^0$
decay to a common CP eigenstate, respectively, and we have used 
$\Gamma_{12}\ll M_{12}$ to introduce the $B$-$\bar B$ mixing phase
$\phi_M.$ If $|\lambda|=1$ also $\bar A/A=e^{-2i\phi_D} $ 
is a pure phase and  the time dependent 
CP asymmetry takes a particularly simple form
\bea
 a_{CP}(t)=-Im\lambda\sin(\Delta Mt)=\sin 2(\phi_M+\phi_D)\sin(\Delta Mt), 
\label{acp}
\eea
where $\Delta M$ is the mass difference between the two physical states.
From \Eq{lambda} and \Eq{acp} it is clear that any new physics effect 
in the mixing will translate into  $\phi_M\to \phi_M+\delta_M$ and will be
 universal to 
all decays while the effect in the decay, $\phi_D\to\phi_D+\delta_D,$
will depend on the process. In the SM the mixing is already one-loop effect 
and therefore  new physics contribution to it may be sizeable.
Without rigourous 
arguments some of the  recent reviews 
\cite{quinn} claim that the 
LRSM contributions
to the $B^0$  mixing are negligible.  We  show the opposite 
by performing an explicit  calculation.

Let us assume that the off-diagonal element $M_{12}$ of the $B_q$-$\bar B_q$
mixing is changed by a factor of $\Delta_q$ as a result of the new 
contribution from the LRSM,
$ M_{12}=M_{12}^{LL}+M_{12}^{LR}=M_{12}^{LL}\Delta_q .$ 
Here $LL$ denotes the contribution from the left-handed sector
which in our convention is equal to the SM result, and $LR$ denotes
the dominant new  contribution from the box diagrams
with one $W_1$ and one $W_2$ and from the tree level flavor changing
Higgs exchange.  $M^{LR}_{12}$ including the LO
 QCD corrections  has been calculated in Ref. \cite{cpb} 
using the vacuum insertion approximation. With $m_b(m_b)=4.4$ GeV,
$m_t(M_1)=170$ GeV, $M_B=5.3$ GeV, $\Lambda^{(5)}_{\overline{MS}}=225$ MeV, 
$\sqrt{B_B}f_B=200$ MeV and the SM input as in Ref. \cite{buras}
the LO QCD improved result reads \cite{cpb} 
\bea
\kappa = F(M_2) \left(\frac{1.6 \mbox{TeV}}
{M_2} \right)^2 +  \left(\frac{12 \mbox{TeV}}
{M_H} \right)^2 \; ,
\label{m12ratio}
\eea
where  $\kappa=|M_{12}^{LR}|/|M_{12}^{LL}|$ and
the function $F(M_2)$ is a complicated function of $W_2.$
Numerically $F(1.6\,\mrm{TeV})=0.2$ 
and $F(10.\,\mrm{TeV})=0.5.$ 
Note  that this estimate holds for both $B^0_d$ and $B^0_s$ systems.
One can write $\Delta_q=1+\kappa e^{i\sigma_q},$
where $\sigma_q = \mbox{Arg}(M_{12}^{LR}/M_{12}^{LL}).$
Consequently the phase $\phi_M^q$ in the mixing in the LRSM becomes
$\phi_M^q=\phi_M^{SM,q}+ \delta_M^q$ where
\bea
\delta_M^q=\arctan\left(\frac{\kappa\sin\sigma_q}
{1+\kappa\cos\sigma_q}\right) \,.
\label{deltam}
\eea 
The phase 
$e^{i\sigma_q} \simeq  - (V_{R,tq} V_{R,tb}^*)/(V_{L,tq} V_{L,tb}^*) $
in our model has been calculated in terms of the quark masses and phase
$\omega$ and reads \cite{cpk}
$\sin\sigma_{d} \simeq  \pm k_2/k_1 \sin\omega $$
[2 \mu_c/\mu_s ( 1 + s_1^2 \mu_s/(2 \mu_d) )
+ \mu_t/\mu_b] ,$
$\sin\sigma_{s} \simeq  \pm k_2/k_1 \sin\omega [ \mu_c/\mu_s $$
 + \mu_t/\mu_b ]  ,$
where $\mu_i = \pm m_i$ and $\pm$ are the signs occuring in the 
Yukawa sector of the model.
While $|k_2/k_1 \sin\omega|\le m_b/m_t$ \cite{cpk}
there is an enhancement factor $m_t/m_b$ in the expressions for  
$\sin\sigma_{d,s}$ which thus can be as large as unity. 
Therefore, taking into account the present constraints on the 
right-handed particle masses  it follows from 
Eqs \rfn{m12ratio}, \rfn{deltam} 
that in the LRSM with spontaneous CP violation 
the phases $\delta_M^q$ can take any value from 0 to $2\pi$ and, consequently, 
the CP asymmetries in \Eq{acp} can vary between -1 and 1.

Unfortunately the CP asymmetries in $B^0_s$ decays which are predicted
to be very small in the SM and can easily show up the new physics 
cannot be studied in $B$ factories
running on the $\Upsilon$ peak. $B^0_d$ decays, however, involve large 
CP asymmetries which are  predicted with poor accuracy in the SM.
The "benchmark" modes $B\to J/\psi K_S$ and $B\to \pi^+\pi^-$ measure
$a_{CP}=\sin 2\beta $ and $a_{CP}=\sin 2\alpha,$ respectively,
where $\beta$ and $\alpha$ are the angles of the SM unitary triangle.
The SM predictions for them are 
$0.3\lsim\sin 2\beta\lsim 0.9 $ and $|\sin\alpha|\le 1$ \cite{quinn}. 
Unless the 
experimental measurement $\beta_{exp}=\beta+\delta_M$ 
clearly lays outside the allowed
region the new physics cannot be traced off. Moreover, since $\alpha$ 
gets modified as $\alpha_{exp}=\alpha-\delta_M$ then
$\delta_M$  cancels out in $\alpha_{exp}+\beta_{exp}$ \cite{nir}. 
Therefore, finding 
new physics could rely only on the experimentally very challenging
measurement of the third angle $\gamma_{exp}.$
 
On the other hand, it is known that in the SM the CP asymmetries 
in the theoretically clean decays  
$B_d\to J/\psi K_S$ ($b\to c\bar c s$) and $B_d\to \phi K_S$ 
($b\to s\bar ss$) measure
with high accuracy the same  angle $\beta.$ The uncertainty in the SM is
estimated to be \cite{gross}
\bea
|\phi(B_d\to J/\psi K_S) - \phi(B_d\to \phi K_S)| \lsim 0.04\,,
\label{phi}
\eea  
where $\phi=\phi_M+\phi_D.$ 
Any deviation from this relation (which should be further tested as proposed
in Ref. \cite{giw}) will be a clear indication of new 
physics. The decay $b\to c\bar c s$
is dominated by tree level $W_1$ exchange and the new physics contribution to
it cannot be sizeable. However, the flavor changing decay
$b\to s\bar ss$  is one-loop
effect in the SM and can, therefore,  be modified by new physics.

The flavor changing decay $b\to s\bar ss$ is induced by 
the  QCD-, electroweak-
and magnetic penguins.  The dominant contribution comes from the QCD penguins
with top quark in the loop.
It is also known \cite{fleish1} that the electroweak penguins 
decrease  about 30\%  the decay rate and
we shall add their contribution to the QCD improved effective Hamiltonian. 
We start with
the effective Hamiltonian due to the gluon exchange
describing the decay $b\to s\bar ss$
at the scale $M_1$  
\bea
H^0_{eff}= -\frac{G_F}{\sqrt{2}}\frac{\alpha_s}{\pi} V_L^{ts*}V_L^{tb}
\left( \bar s\left[\Gamma_\mu^{LL}+\Gamma_\mu^{LR} \right] T^a b\right)
\left( \bar s \gamma^\mu T^a s \right),
\label{heffmw}
\eea
where
$\Gamma_\mu^{LL}=E_0(x_t)\gamma_\mu P_L +
2im_b/q^2E_0'(x_t)\sigma_{\mu\nu}q^\nu P_R, $
$\Gamma_\mu^{LR}=2im_b/q^2\tilde E_0'(x_t)
[A^{tb} \sigma_{\mu\nu}q^\nu P_R +
A^{ts*} \sigma_{\mu\nu}q^\nu P_L ],$
and the $\Gamma_\mu^{LR}$ term describes the new dominant
left-right contribution  via the mixing angle $\xi.$ 
Here $A^{tb}=\xi m_t/m_b V_R^{tb}/V_L^{tb}e^{i\omega}\equiv
\xi m_t/m_b e^{i\sigma_1}$ and analogously
$A^{ts}=\xi m_t/m_b V_R^{ts}/V_L^{ts} e^{i\omega}\equiv
\xi m_t/m_b e^{i\sigma_2}.$ Note that the phases $\sigma_{1,2}$
are independent and can take any value in the range $(0,2\pi).$
 The functions $E_0(x_t),$ $E_0'(x_t)$ and
$\tilde E_0'(x_t)$ are Inami-Lim type functions \cite{lim} of 
$x_t=m_t^2/M_1^2$ and are given by 
$E_0(x_t)=-2/3\ln x + x(18-11x-x^2)/(12(1-x)^3)+
x^2(15-16 x+4x^2)/(6(1-x)^4)\ln x, $
$E_0'(x_t)= x(2+5x-x^2)/(8(x-1)^3)-
3x^2/(4(x-1)^4)\ln x, $
$\tilde E_0'(x_t)= -(4+x+x^2)/(4(x-1)^2)+
3x/(2(x-1)^3)\ln x .$
The left-right analog of
 $E_0'(x_t)$,
$\tilde E_0'(x_t)$, 
is numerically about factor of four larger than the  
latter one. Together with the $m_t/m_b$ enhancement in $A^{tq}$
this practically overcomes the left-right suppression by small $\xi.$

To obtain  reliable estimates for the CP asymmetries 
in $b\to s\bar ss$ induced modes in the LRSM we 
have to calculate the LO QCD corrections to \Eq{heffmw}.
Using the operator product expansion to integrate out the heavy 
fields and  calculating  the  Wilson coefficients $C_i$ in 
the leading logarithm approximation we run them with the renormalization 
group equations from the scale of $W_1$ 
down to the scale $\mu=m_b$ (since the contributions of $W_2, H^0_{1,2}$
are negligible we  start immediately
from the $W_1$ scale).  Because the new physics appears only in the  
gluonic magnetic  operators we can safely take over some well-known
results from the SM studies. 
The effective Hamiltonian we work with is
\bea
H_{eff}=-\frac{G_F}{\sqrt{2}} V_L^{ts*}V_L^{tb}\left(
\sum_{i=1}^{20} C_i(\mu) O_i(\mu) +\sum_{j=7}^{10} C^{ew}_j(\mu) O^{ew}_j(\mu)
\right), \nn
\eea
where we have explicitly separated the electroweak penguin operators
(the second term)
which to a good approximation will not receive any new contribution in the 
LRSM from the twenty operators which do mix with the gluonic and 
photonic magnetic operators. Due to the left-right symmetry the twenty 
operators split into two groups, $O_1$-$O_{10}$ and $O'_1$-$O'_{10},$
 which can be obtained by $P_L\leftrightarrow P_R$ from each other. 
For the QCD penguin operators $O_1$-$O_{6},$ magnetic penguin operators
$O_{7,8}$  as well as for the electroweak 
penguin operators $O^{ew}_7$-$O^{ew}_{10}$
 we use the standard set of the operators
from Ref. \cite{buras}. 
The new left-right operators $O_{9,10}$ are  \cite{misiak}
$O_9= 4(m_b/m_c)(\bar s_\alpha\gamma^{\mu}P_Lc_\beta )
(\bar c_\beta \gamma_\mu P_Rb_\alpha)$ and 
$O_{10}= 4(m_b/m_c)(\bar s_\alpha\gamma^{\mu}P_Lc_\alpha )
(\bar c_\beta \gamma_\mu P_Rb_\beta).$
Keeping only the top and bottom  quark masses
 to be non-vanishing, the matching conditions 
at $W_1$ scale are given as $C_2(M_1)=1,$  
$C_7(M_1)=D_0'(x_t)+A^{tb}\tilde D_0'(x_t),$  
$C_7'(M_1)=A^{ts*}\tilde D_0'(x_t),$  
$C_8(M_1)=E_0'(x_t)+A^{tb}\tilde E_0'(x_t),$  
$C_8'(M_1)=A^{ts*}\tilde E_0'(x_t)$ and the rest of the coefficients vanish.  
Here the SM function $D_0'(x_t)$  and 
its left-right analog $\tilde D_0'(x_t)$ are given by
$D_0'(x_t)= x(7-5x-8x^2)/(24(x-1)^3)-
x^2(2-3x)/(4(x-1)^4)\ln x, $
$\tilde D_0'(x_t)= (-20+31x-5x^2)/(12(x-1)^2)+
x(2-3x)/(2(x-1)^3)\ln x\,. $

The $20\times 20$ anomalous dimension  matrix decomposes into two identical
$10\times 10$ submatrices. The SM $8\times 8$ 
submatrix of the latter one can be found in Ref. \cite{rome} and
the rest of the entries have been calculated by Cho and Misiak in Ref. 
\cite{misiak}.
In the leading logarithm approximation the low energy Wilson 
coefficients for five flavors are given by 
$C_i(\mu=m_b)=\sum_{k,l}(S^{-1})_{ik}(\eta^{3\lambda_k/46})S_{kl} C_l(M_1),$ 
where the $\lambda_k$'s in the exponent of $\eta=\alpha_s(M_1)/\alpha_s(m_b)$
are the eigenvalues of the anomalous dimension matrix over $g^2/16\pi^2$
and the matrix $S$ contains the corresponding eigenvectors. 
We find
\bea
C_8(m_b)&=&\eta^{\frac{14}{23}}(E_0'(x_t)+A^{tb}\tilde E_0'(x_t)) +
\sum_{i=1}^5 h_i \eta^{p_i}\,, \\
C_8'(m_b)&=&\eta^{\frac{14}{23}}A^{ts*}\tilde E_0'(x_t) \,,
\eea
where $h_{i}=(0.8623,$ -0.9135, 0.0209, 0.0873, -0.0571) and
$p_{i}=(14/23,$ 0.4086, 0.1456, -0.4230, -0.8994). 
We reproduced $C_7$ and $C_7'$ exactly as in Ref. \cite{misiak}
and $C_3$-$C_6$ numerically within 1\% as in Ref. \cite{buras}
and we shall not present them here.

Denoting $\langle O \rangle\equiv\langle K_S\phi|O|B \rangle$
the decay amplitude of $B\to K_S\phi$ can be written
\bea
\langle H_{eff} \rangle = -\frac{G_F}{\sqrt{2}}V_L^{tb}V_L^{ts*}
\sum_{3-6, 8, 8',ew} C_i(\mu)\langle O_i(\mu) \rangle\,.
\eea
Contributions from $\langle O_7 \rangle $ are suppressed  
by a factor of $\sqrt{3\alpha_s/\alpha}\approx 9.7$ if compared with 
$\langle O_8\rangle $ and therefore negligible.
The hadronic matrix elements $\langle O_8\rangle $ and 
$\langle O_8'\rangle $ can be approximated to be of the form
\bea
 \langle O_8\rangle =-\frac{2\alpha_s}{\pi}\frac{m_b}{q^2}
\langle \bar s i\sigma_{\mu\nu}q^\mu P_R T^a b 
\bar s \gamma^\nu T^a s\rangle \,,
\eea
and similarly for $ \langle O_8'\rangle ,$ 
where the timelike gluon has produced $\bar ss.$  
Using factorization and the following parametrization for the hadronic matrix
elements \cite{bsw}  
$\langle \phi|\bar s\gamma_\mu s|0 \rangle$=$f_\phi M_\phi \epsilon_\mu,$
$\langle K|\bar s\gamma^\mu b|B \rangle$=$(q_+ -\Delta q_-)F_{BK}(q^2_-;1^-)
+\Delta q_-F'_{BK}(q^2_-;0^+),$ where $f_\phi$=0.23 GeV,
  $q_\pm$=$q_B\pm q_K$ and 
$\Delta$=$(M_B^2-M_K^2)/q_-^2$ one gets  \cite{fleish1}
$\langle O_3 \rangle$=$\langle O_4 \rangle$=$4a/3,$ 
$\langle O_5 \rangle$=$a,$ $\langle O_6 \rangle$=$a/3,$ 
$\langle O_7^{ew} \rangle$=$-a/2,$ $\langle O_8^{ew} \rangle$=$-a/6$ and  
$\langle O_9^{ew} \rangle$=$\langle O_{10}^{ew} \rangle$=$-2a/3,$ 
where  $a$=$f_\phi M_\phi F_{BK}(M_\phi^2;1^-)q_+\cdot\epsilon_\phi.$
In the parametrization of \cite{bsw} 
$F_{BK}(M_\phi^2;l)$=$F_{BK}(0)/(1-M_\phi^2/M_{BK}^2(l)),$ where 
$F_{BK}(0)$=0.38,  $M_{BK}(0^+)$=5.8 GeV and $M_{BK}(1^-)$=5.4 GeV. 
The element $\langle O_8\rangle$ decomposes to
$ \langle O_8\rangle$=$-i4\alpha_s m_b/(9\pi q^2)\langle
\bar s(\gamma^\mu q_s^\nu-
\gamma^\nu q_s^\mu)P_L s\bar s \sigma_{\mu\nu} b-i2m_b\bar s \gamma^\mu P_Ls
\bar s \gamma_\mu P_L b\rangle,$ where $q^\mu_s\approx q_\phi^\mu/2.$ 
With  factorization the new 
matrix element appearing is 
$\langle K|\bar s\sigma^{\mu\nu} b|B\rangle$=$if_\sigma(q_+^\mu q_-^\nu -
q_+^\nu q_-^\mu),$ where $f_\sigma$=$[(1+\Delta)F_{BK}(M_\phi^2;1^-)-
\Delta F_{BK}(M_\phi^2;0^+)]/(4m_b)\approx F_{BK}(M_\phi^2;1^-)/(4m_b)$ 
is obtained using the heavy quark
effective theory \cite{isgur}. As a result we get 
$\langle  O_8 
\rangle$=$-2\alpha_s/(9\pi)m_b^2/q^2 a[1-M^2_\phi/(4m_b^2)],$
where the second term in brackets is negligibly small. The same result 
is valid also for  $\langle  O_8' \rangle.$

It has been shown that the`` physical" range of $q^2$ in  $B\to K_S\phi$ is
$1/4 \lsim q^2/m_b^2 \lsim 1/2$ \cite{hou}. 
To be conservative we use $q^2=m_b^2/2,$ 
$\xi=0.01$ and $m_b(M_1)=2.8$ GeV \cite{pich} to estimate the 
possible effects of new physics.
Numerically we obtain for the LO QCD improved amplitude 
$A\equiv\langle H_{eff}\rangle$
\bea
A=-\frac{G_F}{\sqrt{2}}V_L^{tb}V_L^{ts*}
[-0.0154+0.0047(e^{i\sigma_1}+e^{-i\sigma_2})]a\,.
\label{ares}
\eea 
It is important to notice that in  $B\to \phi K_S$ {\it both} phases
$\sigma_{1,2}$ contribute to the CP asymmetry because only the hadronic
matrix elements of the vector currents matter. This should be compared
with $B\to X_s\gamma$ in which CP asymmetry is given only by the 
right-projected operators \cite{atwood} and, 
consequently,  the phase $\sigma_2$ 
does not contribute.  Also, the major source of uncertainty in the 
decay rate, the hadronic matrix element $a,$ cancels out in the CP
asymmetry. The maximum effect is obtained if 
$\sigma_1=-\sigma_2=\pi/2+\delta_D.$
We get $(\bar A/A)_{max}=e^{1.3 i}$ which implies $|\delta_D|_{max}=0.65.$
This result should be compared with \Eq{phi} which implies that 
there could be a clear effect of the new physics. 
The maximum allowed difference of
the CP asymmetries in $B\to J/\psi K_S$ and $B\to \phi K_S$
in the LRSM could thus be as large as
$|a_{CP}(B\to J/\psi K_S)-a_{CP}(B\to \phi K_S)|_{max}=1.$
If the difference  of the phases in these two processes
will be measured within 10\% and if no difference  will be seen then 
a new upper bound,  $\xi\lsim 0.002,$  can be put
on the left-right mixing angle for {\it large} phases which is
 stronger than the present limit for small phases
$\xi\lsim 0.0025.$

Note  that the new effect in \Eq{ares} is due to the LR contribution to
the QCD magnetic penguins. This new contribution can also provide an answer
to the enhancement of $b\to s g^*$ observed by CLEO \cite{cleo}.

Finally, let us consider the constraints on the LRSM  coming from
the decay $B\to X_s\gamma.$
It is  possible that due to the cancellation between the $LL$ and $LR$
contributions both the total rate $\Gamma$ and 
the CP asymmetry in this process
can, within errors, correspond to the SM predictions \cite{atwood}. 
If the SM predictions will be confirmed experimentally 
(the CP asymmetry in the SM is expected to  be very small)    
this  will constrain the phase $\sigma_1$ and the size of the
$LR$ contribution to $B\to X_s\gamma$ but cannot probe the phase $\sigma_2$
which will still be a free parameter. 
Assuming $\Gamma_{LRSM}(B\to X_s\gamma)/\Gamma_{SM}(B\to X_s\gamma)=1$
we obtain in the most conservative case
for the decay  $B\to \phi K_S$ in the LRSM that 
$(\bar A/A)_{max}=e^{0.47 i}$ which means $|\delta_D|_{max}=0.24$ and 
$|a_{CP}(B\to J/\psi K_S)-a_{CP}(B\to \phi K_S)|_{max}=0.45.$
Therefore, large observable effects are possible independently of
the results in $B\to X_s\gamma.$

In conclusion, we show that the LRSM with spontaneous violation of CP
can dramatically affect the time dependent CP asymmetries in $B^0_{d,s}$
decays. Due to the new contribution to the $B^0$-$\bar B^0$ mixing
the CP asymmetries can vary from -1 to 1 in both  $B^0_{d}$ and $B^0_{s}$
decays. 
Moreover, the $B$-meson decay amplitudes can receive significant 
new contributions as well.
Most importantly for discovering the new physics in the $B$
factories, the CP asymmetries in  
$B\to J/\psi K_S$ and $B\to \phi K_S$ which are equal with high accuracy
 in the SM
can differ from each other as much as unity in our model
independently of the results in  $B\to X_s\gamma.$
Interestingly, the excess of $b\to s g^*$ observed by CLEO 
can also be explained by the LRSM. 

We thank Y. Grossman for pointing out the interesting CP violation effects 
in the decay amplitudes and A. Pich and L. Silvestrini
 for discussions on the QCD corrections.

\end{document}